\documentclass[12pt]{article}
\usepackage{graphicx,amsfonts}

\begin{document}

\title{
{\large \center (Extrait du) {\it Bulletin de l'Acad\'emie royale de Belgique} \\ (Classe des Sciences) \\
$\bf 5^e$ s\'erie - Tome LVIII \quad 1972-1  p. 86-104 \\ }
Canonical  Formalism in Special Relativity}

\author{by E. Pi\~na \thanks{Present\'e par M. R. Balescu} \\
Facult\'e des Sciences Universit\'e Libre de Bruxelles \thanks{Present address: Physics Department, Universidad Aut\'onoma Metropolitana-Iztapalapa, POBox 55 534, M\'exico, D. F. 09340 Mexico \newline
e-mail: pge@xanum.uam.mx} }

\date{S\'eance du samedi 8 janvier 1972}
\maketitle

\abstract{
A covariant Hamiltonian description was introduced in the dynamics of charges and electromagnetic interaction. By a canonical transformation this Hamiltonian formalism was transformed to obtain the Dirac generators for any form of relativistic dynamics, as coefficients of a first degree polynomial in the ten translation and rotation velocities of the Poincar\'e transformation. The Currie's world line conditions  were generalized to any form of the dynamics. The explicit relation between the covariant field variables and the more usual 3-dimensional Fourier variables was derived.
}

\section*{Introduction}
In this paper we will consider different Hamiltonian formulations of relativistic dynamics , to establish the connection between them.

In order to be concrete, we will adopt as our model an ensemble of point charges interacting through an electromagnetic field.

The equations of motion for this system are the Lorentz equations of motion for the particles and the Maxwell equations for the field. As is well known these equations are easily written in Minkowski space with a tensorial notation \cite{pp} showing immediately their invariant character for all the coordinate systems related by Lorentz transformations.

However, for the Quantum Theory and/or the Statistical Mechanics it is convenient to derive these equations of motion from a Hamiltonian theory.

We will begin our paper constructing a Hamiltonian function using only tensorial quantities, making evident the covariant character of the theory. For this purpose we use the generalized Hamiltonian dynamics developed by Dirac \cite{d1}.

On the other hand, it is possible to take into account the covariant aspect of the relativistic theory, working with a set of canonical generators for the Lorentz group \cite{bk}. The origin for this other formalism is also due to Dirac \cite{d2} who constructed several sets of generators, relating each set to a constraint on the particle coordinates. He calls each set $\ll$ a form $\gg$ (of dynamics).

This point of view has been adopted by several authors to study various aspects of relativistic mechanics.

With this formalism many authors beginning with Bakamjian-Thomas \cite{bt} and Foldy \cite{fo} have constructed generators, depending only on the canonical coordinates for the particles. Currie \cite{cu}, \cite{cj} showed that these generators may be consistent with invariant trajectories only for the non interacting case.

Balescu and Kotera \cite{bk} worked the Dirac's instant form of dynamics introducing canonical variables for the field. The non-interaction Currie theorem does not apply for this case and they developed an interesting basis for relativistic statistical mechanics on this basis.

In order to avoid Currie's non-interaction theorem we have always used canonical variables for the field. This selection requires renormalization techniques in order to suppress the singularities in the field. But these difficulties are not fundamental ones \cite{d3}.

These field variables are very useful from the physical point of view, when one wishes to express in simple form many of the electromagnetic phenomena such as radiation, absorption, dispersion, etc.

In this paper we will consider the relation between our tensorial Hamiltonian description and the canonical generators formalism. We will develop a systematic technique to construct the generators for any form of the dynamics. We will generalize the Currie world line condition for any form of the dynamics; and we finally will study the instant form according to our general formalism.

We have used the electromagnetic field only for simplicity, but it is possible to generalize this theory for the Van Dam-Wigner interaction \cite{vw}.

\section{Hamiltonian Dirac's formulation}
Our system will be a collection of particles interacting through a field. The equations of motion are \cite{pp}
$$
m_j g_{\alpha \beta} \frac{d}{d \tau}
\frac{\dot{x}^\beta_j}{[g_{\mu \nu} \dot{x}^\mu_j \dot{x}^\nu_j]^{1/2}} = e_j \dot{x}^\beta_j \left[ \frac{\partial A_\beta(x_j)}{\partial x^\alpha} - \frac{\partial A_\alpha(x_j)}{\partial x_\beta} \right] \eqno(1.1)
$$
and
$$
\square A^\mu(x) = 4 \pi J^\mu(x)\, . \eqno(1.2)
$$

Where $g_{\alpha \beta}$ is the metric tensor
$$
g_{\alpha \beta} = \left( \begin{array}{rrrr}
1 & 0 & 0 & 0 \\
0 & -1 & 0 & 0 \\
0 & 0 & -1 & 0 \\
0 & 0 & 0 & -1
\end{array} \right)\, . \eqno(1.3)
$$

$\square$ is the D'Alambertian
$$
\square = g^{\alpha \beta} \frac{\partial}{\partial x^\alpha} \frac{\partial}{\partial x^\beta}\, . \eqno(1.4)
$$
$J^\mu$ is the 4-vector current density
$$
J^\mu(x) = \sum_{j=1}^N e_j \int_{-\infty}^\infty d \tau \delta^4(x-x_j(\tau)) \frac{x^\mu_j(\tau)}{d \tau}\, ; \eqno(1.5)
$$
where: $m_j$ is the mass of the particle $j$, $e_j$ is the charge of the same particle and $\tau$ is an arbitrary parameter used to parametrize the trajectories of the particles. The dot denotes the derivative with respect to the parameter $\tau$.

Further, $x^\mu_j$ is the $\mu$-component of the particle $j$ and $A^\beta(x)$ is the $\beta$-component of the 4-vector electromagnetic potential  at the point $x^\mu$ in space-time.

We look for a formal solution of eq. (1.2). Let us introduce the Fourier transform $q^\mu_\kappa$ of the 4-vector potential

$$
A^\mu(x) = \int d^4 \kappa \frac{1}{\sqrt{4 \pi^3 \kappa_\alpha \kappa^\alpha}} q^\mu_\kappa \exp[i \kappa_\beta x^\beta]
\, . \eqno(1.6)
$$
It follows the formal solution of eq. (1.2)
$$
q^\mu_\kappa = - \frac{1}{\sqrt{4 \pi^3 \kappa_\alpha \kappa^\alpha}} \sum_{j=1}^N e_j \int\limits_{-\infty}^\infty d \tau \dot{x}^\mu_j(\tau) \exp[-i \kappa_\alpha x^\beta_j(\tau)] \, .\eqno(1.7)
$$

We define the canonical variables for the field, the coordinates
$$
y^\mu_\kappa = - \frac{1}{\sqrt{4 \pi^3 \kappa_\alpha \kappa^\alpha}} \sum_{j=1}^N e_j \int\limits_{-\infty}^\tau d \tau \dot{x}^\mu_j(\tau) \exp[-i \kappa_\alpha x^\beta_j(\tau)] \, \eqno(1.8)
$$
and the canonical momenta
$$
p^\mu_\kappa = - \frac{1}{\sqrt{4 \pi^3 \kappa_\alpha \kappa^\alpha}} \sum_{j=1}^N e_j \int\limits_\tau^\infty d \tau \dot{x}^\mu_j(\tau) \exp[i \kappa_\alpha x^\beta_j(\tau)] \, . \eqno(1.9)
$$

They are related to the Fourier transform of the 4-vector potential by the expression
$$
q^\mu_\kappa = y^\mu_\kappa + p^\mu_{-\kappa}\, . \eqno(1.10)
$$

We note that $q^\mu_\kappa$ is a constant of motion.

From the definitions (1.8) and (1.9) we find the equations of motion for the field variables
$$
\dot{y}^\mu_\kappa = - \frac{1}{\sqrt{4 \pi^3 \kappa_\alpha \kappa^\alpha}} \sum_{j=1}^N e_j \dot{x}^\mu_j(\tau) \exp [-i \kappa_\beta x^\beta_j(\tau)] \eqno(1.11)
$$
and
$$
\dot{p}^\mu_\kappa = \frac{1}{\sqrt{4 \pi^3 \kappa_\alpha \kappa^\alpha}} \sum_{j=1}^N e_j \dot{x}^\mu_j(\tau) \exp [i \kappa_\beta x^\beta_j(\tau)] \, . \eqno(1.12)
$$

On the other hand the 4-vector potential may be written as a function of the field variables as follows
$$
A^\mu(x) = \int d^4\kappa \frac{1}{\sqrt{4 \pi^3 \kappa_\alpha \kappa^\alpha}} \{ y^\mu_\kappa \exp[i \kappa_\beta x^\beta] + p^\mu_\kappa \exp[-i \kappa_\beta x^\beta]\} \, . \eqno(1.13)
$$

We are going to consider Dirac's formalism \cite{d1} introduced to generalize the Hamiltonian formulation of dynamics; it is particularly useful for the case where the Lagrangian is a first order homogeneous function of the velocities.

For this situation the ordinary Hamiltonian is identically equal to zero and one finds a family of constraints among the canonical variables
$$
\phi_n(x, p) = 0 \quad \mbox{(n=1, 2, ...)}\, . \eqno(1.14)
$$

Dirac introduces then the generalized Hamiltonian
$$
H = \sum_n v_n \phi_n(x, p) \, , \eqno(1.15)
$$
where the $v_n$'s are Lagrange multipliers.

We will work out a formalism of this type for the simple situation in which the functions $\phi_n$ that determines the constraints of the system satisfy the restriction that the Poisson bracket between any couple of them is equal to zero
$$
[\phi_n, \phi_m] = 0 \, . \eqno(1.16)
$$

Let us introduce the constraints (one for each particle
$$
\phi_j = [p^\alpha_j - e_j A^\alpha(x_j)] g_{\alpha \beta} [p^\beta_j - e_j A^\beta(x_j)] - m_j^2 = 0\, .\eqno(1.17)
$$
Dirac's Hamiltonian now becomes
$$
H = \sum_{j=1}^N v_j \{ p^\alpha_j - e_j A^\alpha(x_j)] g_{\alpha \beta} [p^\beta_j - e_j A^\beta(x_j)] - m_j^2\} \, , \eqno(1.18)
$$
where $A^\mu(x)$ is given explicitly in terms of canonical field variables as in (1.13). From Hamilton equations associated to the Hamiltonian (1.18) and making use of the constraints (1.17) it is possible to determine the $v_j$ as follows
$$
v_j = \frac{1}{2 m_j} [g_{\alpha \beta} \dot{x}^\alpha_j \dot{x}^\beta_j]^{1/2} \, . \eqno(1.19)
$$

This means that the Lagrange multipliers $v_j$ are proportional to the $\ll$ velocity $\gg$ along the world line of the respective particle, measured in $\tau$-units.

Substituting (1.19) into the Hamilton equations for particles and field variables we recover the equations of motion (1.1), (1.11) and (1.12) for particles and field.

In order to arrive to the field equation (1.2) we must to use the boundary conditions
$$
\lim\limits_{\tau\to-\infty} y^\mu_\kappa(\tau) = 0 \eqno(1.20)
$$
and
$$
\lim\limits_{\tau\to\infty} p^\mu_\kappa(\tau) = 0 \, . \eqno(1.21)
$$

It is interesting to note that eqs. of motion are invariants with respect to a change of parameter $\tau$
$$
\tau \longrightarrow F(\tau) \, . \eqno(1.22)
$$

This property implies an undetermined character of the eqs. of motion, as long as the $\tau$ parameter is not fixed by additional restrictions.

In the next section we will introduce a different dynamical approach to give a determined aspect to the equations of motion.

\section{Dirac's canonical formulation}
In this section we want to relate the previous formulation to another one associated to Dirac's ideas.

Dirac set up \cite{d2} a canonical representation for the Poincar\'e group
$$
[P_\mu, P_\nu] = 0\, , \eqno(2.1)
$$
$$
[M_{\mu \nu}, P_\lambda] = g_{\lambda \nu} P_\mu - g_{\lambda \mu} P_\nu\, , \eqno(2.2)
$$
$$
[M_{\alpha \beta}, M_{\mu \nu}] = g_{\alpha \nu} M_{\beta \mu} + g_{\beta \mu} M_{\alpha \nu} + g_{\alpha \mu} M_{\nu \beta} + g_{\beta \nu} M_{\mu \alpha}\, . \eqno(2.3)
$$

Starting from a trivial (geometrical) representation for this group, he proposes to construct a new one, where the ten generators $P_\nu$, $M_{\alpha \beta}$, must obey the Lie algebra restrictions of this group, using as Lie bracket the canonical Poisson bracket. He also demands the ten generators to have zero Poisson bracket with a particular function of the coordinates which specifies the $\ll$ form $\gg$ of the dynamics. He gave some solutions, but he did not introduce any specific technique to find these solutions. For instance Dirac considers:

The instant form
$$
\mathfrak{q}^0 = 0 \, . \eqno(2.4)
$$

The light cone form
$$
\mathfrak{q}^\alpha \mathfrak{q}_\alpha = 0 \, . \eqno(2.5)
$$

The hyperboloid form
$$
\mathfrak{q}^\alpha \mathfrak{q}_\alpha - A^2 = 0 \, . \eqno(2.6)
$$

Etc.

Our main aim will be to give a more clear physical or geometrical meaning to this formalism, to obtain it systematically from the Hamiltonian formalism developed in the previous section, and to give a synthetic method for obtaining the solution to the Dirac's problem of constructing a canonical representation of the Poincar\'e group, consistent with any $\ll$ form $\gg$ of the dynamics.

We consider the $\tau$-dependent Lorentz inhomogeneous transformation obtained by canonical transformation of the Hamiltonian problem presented in last section. This canonical transformation will be generated by the function \cite{go}
$$
F_2 = \sum_{j=1}^N \mathfrak{p}_{j \nu} [a^\nu_{\: \mu} x^\mu_j - z^\nu] + \int d^4 k \mathcal{P}_{k \beta} a^\beta_{\; \mu} y^\mu_\kappa \exp[i k_\gamma z^\gamma] \, , \eqno(2.7)
$$
where: $a^\nu_{\; \mu}$ are the components of the Lorentz transformation matrix; $z^\nu$ is a 4-vector translation; $\mathfrak{p}_{j \nu}$ are the new four momenta of the particles; $\mathcal{P}_{k \beta}$ are the new canonical momenta of the field, and $k^\alpha$ is a new wave vector related to the old one $\kappa^\alpha$ by the same Lorentz transformation
$$
\kappa_\alpha = a^\nu_{\; \alpha} k_\nu\, . \eqno(2.8)
$$

Both quantities $a^\nu_{\; \mu}$ and $z^\nu$ will be considered to be explicit functions of the parameter $\tau$ used to describe the motion. This gives a $\tau$-dependence for the $F_2$ generating function.

Let us make use of $F_2$ to generate the canonical transformation.

The new particle coordinates are
$$
\mathfrak{q}^\nu_j = a^\nu_{\; \mu} x^\mu_j - z^\nu = \frac{\partial F_2}{\partial \mathfrak{p}_{j \nu}} \, , \eqno(2.9)
$$
which is a $\tau$-dependent Poincar\'e transformation among the coordinates of the particles.

Analogously we find the old momenta
$$
p_{j \mu} = \mathfrak{p}_{j \nu} a^\nu_{\; \mu} = \frac{\partial F_2}{\partial x^\mu_j}\, . \eqno(2.10)
$$

The new coordinates for the field are
$$
\mathcal{Q}^\beta_k = a^\beta_{\; \mu} y^\mu_\kappa \exp[i k_\gamma z^\gamma] = \frac{\delta F_2}{\delta \mathcal{P}_{k \beta}} \, \eqno(2.11)
$$

And the old momenta for the field are given by
$$
p_{\kappa \mu} = \mathcal{P}_{k \beta} a^\beta_{\; \mu} \exp[i k_\gamma z^\gamma] = \frac{\delta F_2}{\delta y^\mu_\kappa} \, . \eqno(2.12)
$$

The new Hamiltonian is found by the prescription \cite{go}
$$
\mathcal{H} = H + \frac{\partial F_2}{\partial \tau}\, . \eqno(2.13)
$$

In order to calculate this expression we need the result that follows from (2.11)
$$
\frac{\partial \mathcal{Q}^\beta_k}{\partial k_\nu} = i z^\nu \mathcal{Q}^\beta_k + a^\beta_{\; \mu} \frac{\partial y^\mu_\kappa}{\partial \kappa^\alpha} a^{\nu \alpha} \exp[i k_\gamma z^\gamma] \, . \eqno(2.14)
$$

The $\tau$-derivative of the Lorentz tensor $a^\beta_{\; \mu}$ will be expressed in terms of an antisymmetric tensor as is similarly made in the theory of the rigid rotating body \cite{cs}
$$
\dot{a}^\beta_{\; \mu} = \omega^\beta_{\; \gamma} a^\gamma_{\; \mu} \, , \eqno(2.15)
$$
where $\omega_{\alpha \beta}$ is an antisymmetric angular velocity tensor.

Let us now to calculate the derivative $\partial F_2/\partial \tau$ and afterwards transform it to the new variables by using the equations (2.15), (2.9), (2.11) and (2.14), and the antisymmetric character of the tensor $\omega_{\alpha \beta}$
$$
\frac{\partial F_2}{\partial \tau} = - \dot{z}^\nu \sum_{j=1}^N \mathfrak{p}_{j \nu} + \dot{z}^\nu i \int d^4k k_\nu \mathcal{P}_{k \beta} \mathcal{Q}^\beta_k +
$$
$$
\frac{1}{2} \omega_{\alpha \beta} \int d^4 k\left[ (\mathcal{P}^\alpha_k \mathcal{Q}^\beta_k - \mathcal{P}^\beta_k \mathcal{Q}^\alpha_k ) + \mathcal{P}_{k \gamma} \left(k^\alpha \frac{\partial}{\partial k_\beta} - k^\beta \frac{\partial}{\partial k_\alpha} \right) \mathcal{Q}^\gamma_k\right] +
$$
$$
\frac{1}{2} \omega_{\alpha \beta} i \int d^4k \mathcal{P}_{k \gamma} (k^\alpha z^\beta - k^\beta z^\alpha) \mathcal{Q}^\gamma_k \, . \eqno(2.16)
$$

In order to get the new Hamiltonian as a function of the new variables, we transform the 4-vector potential at the position of particle $j$
$$
\mathcal{A}^\beta(\mathfrak{q}_j) = \int d^4k \frac{1}{\sqrt{4 \pi^3 k_\nu k^\nu}} \{ \mathcal{Q}^\beta_k \exp[i k_\gamma \mathfrak{q}^\gamma_j] + \mathcal{P}^\beta_k \exp[-i k_\gamma \mathfrak{q}\gamma_j]\}\, . \eqno(2.17)
$$

It follows
$$
\mathcal{A}^\beta(\mathfrak{q}_j) = a^\beta_{\; \mu} A^\mu(x_j)\, . \eqno(2.18)
$$

With this result, the old Hamiltonian in the new variables has the same formal aspect as in the previous formulation
$$
H = \sum_{j=1}^N v_j  \{ [\mathfrak{p}^\beta_j - e_j \mathcal{A}^\beta(\mathfrak{q}_j)] g_{\beta \gamma} [\mathfrak{p}^\gamma_j - e_j \mathcal{A}^\gamma(\mathfrak{q}_j)] - m_j^2\}\, .\eqno(2.19)
$$

And the new Hamiltonian is found by adding (2.16) and (2.19) according to (2.13).

The Hamiltonian formulation is completed by taking into account the transformed constraints
$$
[\mathfrak{p}^\alpha_j - e_j \mathcal{A}^\alpha(\mathfrak{q}_j)] g_{\alpha \beta} [\mathfrak{p}^\beta_j - e_j \mathcal{A}^\beta(\mathfrak{q}_j)] - m_j^2 = 0 \, .\eqno(2.20)
$$

We are going now to determine the $\tau$-parametrization by imposing the new constraint for each particle; these constraints fix the $\ll$ form $\gg$ of the dynamics
$$
g(\mathfrak{q}^\alpha_j) = 0 \, , \eqno(2.21)
$$
where $g$ is a point function in the $\mathfrak{q}$ coordinate space.

This constraint in terms of the old coordinates is
$$
g(a^\alpha_{\; \mu} x^\mu_j - z^\alpha) = 0 \, , \eqno(2.22)
$$
that shows more clearly the physical meaning: this constraint fixes the parametrization of the particles by the intersection of the world line of each particle with the $\tau$-dependent family of surfaces
$$
g(a^\alpha_{\; \mu}(\tau) x^\mu - z^\alpha(\tau)) = 0 \, . \eqno(2.23)
$$

Dirac's examples \cite{d2} are now interpreted as follows. In the instant form
$$
g(\mathfrak{q}^\alpha_j) \equiv \mathfrak{q}^0_j = 0 \, , \eqno(2.24)
$$
the particles are parametrized by a family of hyperplanes in original space.

In the light-cone form
$$
g(\mathfrak{q}^\alpha_j) \equiv \mathfrak{q}^\alpha_j g_{\alpha \beta} \mathfrak{q}\beta_j = 0 \, ,
\eqno(2.25)
$$
the particles are parametrized by a family of light-cones.

Etc.

Returning to a general constraint we look now to the preservation, following the motion, of the constraint
$$
0 = \frac{d g}{d \tau} = \frac{\partial g}{\partial\mathfrak{q}^\alpha_j} \dot{\mathfrak{q}}^\alpha_j = [g, \mathcal{H}] \, . \eqno(2.26)
$$

This equation determines the $v_j$ Lagrange multipliers in $\mathcal{H}$ as follows
$$
v_j = \frac{\dot{z}^\beta \frac{\partial g}{\partial \mathfrak{q}^\beta_j} + \frac{1}{2} \omega_{\beta \gamma} [(\mathfrak{q}^\beta_j + z^\beta) g^{\gamma \mu} - (\mathfrak{q}^\gamma_j + z^\gamma) g^{\beta \mu}] \frac{\partial g}{\partial \mathfrak{q}^\mu_j} }{2[\mathfrak{p}^\alpha_j - e_j \mathcal{A}^\alpha(\mathfrak{q}_j)] \frac{\partial g}{\partial \mathfrak{q}^\alpha_j}} \, . \eqno(2.27)
$$

The Hamiltonian $\mathcal{H}$ is now written in Dirac's form \cite{d3}
$$
\mathcal{H} = - \dot{z}^\nu_R P_\nu - \frac{1}{2} \omega^{\alpha \beta} M_{\alpha \beta} \, , \eqno(2.28)
$$
where
$$
\dot{z}^\nu_R = \dot{z}^\alpha - \omega^\alpha_{\; \beta} z^\beta = a^\alpha_{\; \beta} \frac{d}{d \tau} (a_\gamma^{\; \beta} z^\gamma) \, , \eqno(2.29)
$$
$$
P_\nu = \sum_{j=i}^N \mathfrak{p}_{j \nu} - i \int d^4k k_\nu \mathcal{P}_{k \beta} \mathcal{Q}^\beta_k
$$
$$
- \sum_{j=1}^N \frac{\frac{\partial g}{\partial \mathfrak{q}^\nu_j}}{2[\mathfrak{p}^\alpha_j - e_j \mathcal{A}^\alpha(\mathfrak{q}_j)] \frac{\partial g}{\partial \mathfrak{q}^\alpha_j}} \{ [\mathfrak{p}^\beta_j - e_j \mathcal{A}^\beta(\mathfrak{q}_j)] g_{\beta \gamma} [\mathfrak{p}^\gamma_j - e_j \mathcal{A}^\gamma(\mathfrak{q}_j)] - m_j^2\}\, ,\eqno(2.30)
$$
and where $M^{\alpha \beta}$ is the antisymmetric tensor
$$
M^{\alpha \beta} = \sum_{j=1}^N (\mathfrak{q}^\alpha_j \mathfrak{p}^\beta_j - \mathfrak{q}^\beta_j \mathfrak{p}^\alpha_j) +
$$
$$
+ \int d^4k \left[ \mathcal{P}^\beta_k \mathcal{Q}^\alpha_k - \mathcal{P}^\alpha_k \mathcal{Q}^\beta_k + \mathcal{P}_{k \gamma} \left( k^\beta \frac{\partial}{\partial k_\alpha} - k^\alpha \frac{\partial}{\partial k_\beta} \right) \mathcal{Q}^\gamma_k \right]
$$
$$
- \sum_{j=1}^N \frac{ (\mathfrak{q}^\alpha_j g^{\beta \gamma} -\mathfrak{q}^\beta_j g^{\alpha \gamma}) \frac{\partial g}{\partial \mathfrak{q}^\gamma_j}}{2[\mathfrak{p}^\alpha_j - e_j \mathcal{A}^\alpha(\mathfrak{q}_j)] \frac{\partial g}{\partial \mathfrak{q}^\alpha_j}} \{ [\mathfrak{p}^\mu_j - e_j \mathcal{A}^\mu(\mathfrak{q}_j)] g_{\mu \nu} [\mathfrak{p}^\nu_j - e_j \mathcal{A}^\nu(\mathfrak{q}_j)] - m_j^2\}\, .\eqno(2.31)
$$

It is possible to consider here the hamiltonian $\mathcal{H}$ as a Routh function \cite{go}: i.e. as a Lagrangian with respect to the variables $z^\alpha$ and $a^\alpha_{\; \beta}$.

The Lagrangian equation associated with the $z^\alpha$ variable give us
$$
\dot{P}_\mu = \omega^\beta_{\; \mu} P_\beta \, , \eqno(2.32)
$$
which expresses the conservation of the momentum 4-vector
$$
a^\beta_{\; \alpha} P_\beta \, . \eqno(2.33)
$$

Taking into account the Lorentz constraints
$$
a^\beta_{\; \alpha} a_\gamma^{\; \alpha} = \delta^\beta_\gamma \, , \eqno(2.34)
$$
the Lagrange equation associated to the variable $a^\alpha_{\; \beta}$ give us
$$
\frac{d}{d \tau}(M^{\mu \beta} + z^\mu P^\beta - z^\beta P^\mu) =
$$
$$
= \omega^\beta_{\; \alpha} (M^{\mu \alpha} + z^\mu P^\alpha - z^\alpha P^\mu) - \omega^\mu_{\; \alpha} (M^{\beta \alpha} + z^\beta P^\alpha - z^\alpha P^\beta) \, , \eqno(2.35)
$$
which represent the conservation of the antisymmetric angular momentum tensor
$$
a^\mu_{\; \alpha} a^\nu_{\; \beta} (M_{\mu \nu} + z_\mu P_\nu - z_\nu P_\mu) \, . \eqno(2.36)
$$

On the other hand, the Hamiltonian equation of motion for $P_\mu$ is
$$
\dot{P}_\mu = [\mathcal{H}, P_\mu] = - \dot{z}^\nu_R [P_\nu, P_\mu] - \frac{1}{2} \omega^{\alpha \beta} [M_{\alpha \beta}, P_\mu] \, .\eqno(2.37)
$$
Comparing (2.32) and (2.37) it follows that
$$
[P_\nu, P_\mu] = 0
$$
and
$$
[M_{\alpha \beta}, P_\mu] = g_{\beta \mu} P_\alpha - g_{\alpha \mu} P_\beta \, . \eqno(2.39)
$$

Studying the Hamiltonian motion equation for $M_{\mu \nu}$ and comparing with (2.35) we also found
$$
[M_{\alpha \beta}, M_{\mu \nu}] = g_{\alpha \nu} M_{\beta \mu} + g_{\beta \nu} M_{\mu \alpha} + g_{\alpha \mu} M_{\nu \beta} + g_{\beta \mu} M_{\alpha \nu}\, . \eqno(2.40)
$$

These three equations are the fundamental Lie algebra commutators of the Poincar\'e group. We found them as compatibility coditions between the Lagrangian and the Hamiltonian formulations associated to the Routhian $\mathcal{H}$.

The Hamiltonian expression for the preservation of the $g$ constraint
$$
[\mathcal{H}, g] = 0 \, , \eqno(2.41)
$$
implies now the properties
$$
[P_\nu, g] = 0 \eqno(2.42)
$$
and
$$
[M_{\alpha \beta}, g] = 0 \, . \eqno(2.43)
$$

In Dirac's paper \cite{d2} these equations are the starting point for the determination of the ten generators $P_\nu$, $M_{\alpha \beta}$, by an inductive method, different for each constraint. In this paper, on the contrary, the general expression for the generators (2.30) and (2.31) are obtained directly from the tensorial Hamiltonian formulation by applying a canonical transformation to a moving reference frame and finding the $v_j$ Lagrange multipliers with the aid of the constraints $g$ that determine the parametrization.

The method here presented has therefore the double advantage of showing explicitly the connection between the two Dirac formulations and of giving the general expression for the generators valid for any $\ll$ form $\gg$ of the dynamics.

At first sight there is a pathological case in Dirac's paper. The two constraints
$$
g(\mathfrak{q}) = \mathfrak{q}^\alpha \mathfrak{q}_\alpha\quad \mbox{ and }\quad g(\mathfrak{q}) = \mathfrak{q}^\alpha \mathfrak{q}_\alpha - A^2 \eqno(2.44)
$$
should have equal generators in our formulation and however Dirac gives different types of generators in the two cases.

The paradox is solved by noting that the generators $P_\nu$ for Dirac's hyperboloid form may be transformed to the other $P_\gamma$ generators by adding the $\ll$ strong equation $\gg$ (in Dirac's terminology \cite{d1} )
$$
\left\{ \frac{A^2}{2 \mathfrak{p}_\nu \mathfrak{q}^\nu} (\mathfrak{p}_\sigma \mathfrak{p}^\sigma - m^2) \right\}^2 = 0 \, . \eqno(2.45)
$$

In order to verify directly the two equations (2.42) and (2.43) we found the interesting results
$$
[\mathfrak{q}^\mu_j, P^\gamma] = - g^{\mu \gamma} + \frac{g^{\gamma \nu} \frac{\partial g}{\partial \mathfrak{q}^\nu_j} [\mathfrak{p}^\mu_j - e_j \mathcal{A}^\mu(\mathfrak{q}_j) ] }{[\mathfrak{p}^\alpha_j - e_j \mathcal{A}^\alpha(\mathfrak{q}_j)] \frac{\partial g}{\partial \mathfrak{q}^\alpha_j}} \eqno(2.46)
$$
and
$$
[\mathfrak{q}^\mu_j, M^{\alpha \beta}] = (\mathfrak{q}^\alpha_j \delta^\beta_\gamma - \mathfrak{q}^\beta_j \delta^\alpha_\gamma) [\mathfrak{q}^\mu_j, P^\gamma] \, . \eqno(2.47)
$$

These expressions are equivalent to Currie's \cite{cu} conditions for the trajectories of particles. However it is necessary to remark that Currie's original formulae are related to Dirac's instant form studied in the next section, whereas equations (2.46) and (2.47) are valid for an arbitrary $g$-constraint. These equations will guarantee the condition of invariant trajectories of particles, independently of the $\ll$ form $\gg$ selected for the dynamical description.

\section{The instant form}
We will consider in this section the more usual, relativistic form of dynamics related to the Dirac's instant form, where the $g$-constraint is
$$
g(\mathfrak{q}^\alpha_j) \equiv \mathfrak{q}^0_j = 0 \, . \eqno(3.1)
$$

This $\ll$ instant form $\gg$ is specially important because of its physical clearness and its analogy with the non relativistic case.

The constraint (3.1) implies therefore
$$
\frac{\partial g}{\partial\mathfrak{q}^\alpha_j} = \delta^0_\alpha \, . \eqno(3.2)
$$
And as a consequence the generators will take the following form
$$
P^\alpha = \sum_{j=1}^N \mathfrak{p}^\alpha_j - i \int d^4k k^\alpha \mathcal{P}_{k \beta} \mathcal{Q}^\beta_k \quad \mbox{ ($\alpha$ = 1, 2, 3) } \, , \eqno(3.3)
$$
$$
P^0 = \sum_{j=1}^N \mathfrak{p}^0_j -i \int d^4k k^0 \mathcal{P}_{k \beta} \mathcal{Q}^\beta_k
$$
$$
- \sum_{j=1}^N \frac{1}{2[\mathfrak{p}^0_j - e_j \mathcal{A}^0(\mathfrak{q}_j)]} \{ [\mathfrak{p}^\beta_j - e_j \mathcal{A}^\beta(\mathfrak{q}_j)] g_{\beta \gamma} [\mathfrak{p}^\gamma_j - e_j \mathcal{A}^\gamma(\mathfrak{q}_j)] - m_j^2\}\, ,\eqno(3.4)
$$
$$
M^{\alpha \beta} = \sum_{j=1}^N(\mathfrak{q}^\alpha_j \mathfrak{p}^\beta_j - \mathfrak{q}^\beta_j \mathfrak{p}^\alpha_j) + \int d^4k \left\{ \mathcal{Q}^\alpha_k \mathcal{P}^\beta_k - \mathcal{Q}^\beta_k \mathcal{P}^\alpha_k + \mathcal{P}_{k \gamma} \left( k^\beta \frac{\partial}{\partial k_\alpha} - k^\alpha \frac{\partial}{\partial k_\beta}\right) \mathcal{Q}^\gamma_k \right\}
$$
$$
(\alpha \neq \beta = 1, 2, 3) \, , \eqno(3.5)
$$
$$
M^{\alpha 0} = \sum_{j=1}^N \mathfrak{q}^\alpha_j \mathfrak{p}^0_j + \int d^4k \left\{ \mathcal{Q}^\alpha_k \mathcal{P}^0_k - \mathcal{Q}^0_k \mathcal{P}^\alpha_k + \mathcal{P}_{k \gamma} \left( k^0 \frac{\partial}{\partial k_\alpha} - k^\alpha \frac{\partial}{\partial k_0}\right) \mathcal{Q}^\gamma_k \right\}
$$
$$
- \sum_{j=1}^N \frac{\mathfrak{q}^\alpha_j}{2[\mathfrak{p}^0_j - e_j \mathcal{A}^0(\mathfrak{q}_j)]} \{ [\mathfrak{p}^\beta_j - e_j \mathcal{A}^\beta(\mathfrak{q}_j)] g_{\beta \gamma} [\mathfrak{p}^\gamma_j - e_j \mathcal{A}^\gamma(\mathfrak{q}_j)] - m_j^2\}\,
$$
$$
( \alpha = 1, 2, 3) \, . \eqno(3.6)
$$

For positive energy the constraints will take the form
$$
\mathfrak{p}^0_j = e_j \mathcal{A}^0(\mathfrak{q}_j) + \sqrt{[{\bf p}_j - e_j {\bf A}(\mathfrak{q}_j)]^2 + m_j^2} \, , \eqno(3.7)
$$
where we introduced the the 3-vectorial notation
$$
{\bf p}_j = (\mathfrak{p}^1_j, \mathfrak{p}^2_j, \mathfrak{p}^3_j) \eqno(3.8)
$$
and
$$
{\bf A}(\mathfrak{q}_j) = (\mathcal{A}^1(\mathfrak{q}_j), \mathcal{A}^2(\mathfrak{q}_j), \mathcal{A}^3(\mathfrak{q}_j)) \, . \eqno(3.9)
$$

Because $\mathfrak{q}^0_j$ is a constant, according to (3.1), we suppress explicitly the canonical conjugate variable $\mathfrak{p}^0_j$ by using the constraint equation (3.7)

The generators $P^0$ and $M^{\alpha 0}$ are modified to the new expressions
$$
P^0 = \sum_{j=1}^N \left\{ e_j \mathcal{A}^0(\mathfrak{q}_j) + \sqrt{[{\bf p}_j - e_j {\bf A}(\mathfrak{q}_j)]^2 + m_j^2}\right\} - i \int d^4k k^0 \mathcal{P}_{k \beta} \mathcal{Q}^\beta_k \eqno(3.10)
$$
and
$$
M^{\alpha 0} = \sum_{j=1}^N \mathfrak{q}^\alpha_j \left\{ e_j \mathcal{A}^0(\mathfrak{q}_j) + \sqrt{[{\bf p}_j - e_j {\bf A}(\mathfrak{q}_j)]^2 + m_j^2}\right\} +
$$
$$
+ \int d^4k \left\{ \mathcal{Q}^\alpha_k \mathcal{P}^0_k - \mathcal{Q}^0_k \mathcal{P}^\alpha_k + \mathcal{P}_{k \gamma} \left( k^0 \frac{\partial}{\partial k_\alpha} - k^\alpha \frac{\partial}{\partial k_0}\right) \mathcal{Q}^{\gamma}_k \right\}
$$
$$
(\alpha = 1, 2, 3) \, . \eqno(3.11)
$$

For this particular choice of $g$-constraint the equations (2.46) and (2.47) will give us
$$
[\mathfrak{q}^\alpha_j, P^\beta] = \delta^{\alpha \beta} \, ,
$$
$$
[\mathfrak{q}^\gamma_j, M^{\alpha \beta}] = \mathfrak{q}^\beta_j \delta^{\alpha \gamma} - \mathfrak{q}^\alpha_j \delta^{\beta \gamma} \, , \eqno(3.12)
$$
$$
[\mathfrak{q}^\gamma_j, M^{\alpha 0}] = \mathfrak{q}^\alpha_j [\mathfrak{q}^\gamma_j, P^0]
$$
$$
(\alpha = 1, 2, 3; \beta = 1, 2, 3; \gamma = 1, 2, 3) \, .
$$
These are the conditions obtained by Currie \cite{cu}, also Currie, Jordan and Sudarshan \cite{cj}, for the invariance of the trajectories of particles. These formulae are valid only for the formulation that admits the $g$-constraint (3.1). We found previously the equations valid for an arbitrary constraint: they are eqs. (2.46) and (2.47). Eqs. (3.12) were obtained by Currie as compatibility conditions between the Lorentz transformation of the Hamiltonian formulation and the geometrodynamical transformation of the simultaneous positions of the particles. The method is based on comparing infinitesimal transformations with both techniques. This calculation was made with the explicit hypothesis that the coordinates of particles are considered at the same time.

These authors did not remark that these conditions are modified in the case when a different parametrization is used.

If we take the variables $a^\alpha_{\; \beta}$ constants so that
$$
\omega _{\alpha \beta} = 0 \eqno(3.13)
$$
and the variables $z^\mu$ by the equations
$$
z^0 = -\tau \, , \quad z^\alpha = 0 \quad (\alpha= 1, 2, 3). \eqno(3.14)
$$
This choice corresponds to a parametrization by the time measured in an arbitrary frame specified by the constants $a^\alpha_{\; \beta}$.

Introducing (3.13) and (3.14) in the Routhian (2.28) we found the remarkable property
$$
\mathcal{H} = P^0 \, . \eqno(3.15)
$$

Lastly, we would like to point out the relation between the canonical variables for the field used in this paper and the formulation employed currently in the literature \cite{bk}, \cite{he}.

Because of the constraint (3.1) the quantities appearing in the Fourier expression for the 4-vector potential
$$
\mathcal{A}^\beta(\mathfrak{q}_j) = \int d^4k \frac{1}{\sqrt{4 \pi^3 k_\gamma k^\gamma}} [\mathcal{Q}^\beta_k + \mathcal{P}^\beta_{-k}] \exp[i k_\alpha \mathfrak{q}^\alpha_j] \, , \eqno(3.16)
$$
must have a singular character. This remark enables us to diminish the number of dimensions of the functional dependence of the field variables.

We want to obtain the usual expression for the 4-vector potential
$$
\mathcal{A}^\beta(\mathfrak{q}_j) = \int \frac{d^3k}{k} \{ A^\beta_k \exp[-i {\bf k} \cdot {\bf q}_j] + A^{\dagger \beta}_k \exp[i {\bf k} \cdot {\bf q}_j]\} \, , \eqno(3.17)
$$
where
$$
{\bf q}_j = (\mathfrak{q}^1_j, \mathfrak{q}^2_j, \mathfrak{q}^3_j) \, , \eqno(3.18)
$$
$$
{\bf k} = (k^1,  k^2,  k^3) \, , \eqno(3.19)
$$
and where $A^\beta_k$, $A^{\dagger \beta}_k$, are the new field variables. Between these variables the Poisson brackets are given by
$$
[A^\mu_k, A^{\dagger \nu}_{k'}] = - \frac{i}{4 \pi^2} g^{\mu \nu} \delta^3({\bf k} - {\bf k}') \, , \eqno(3.20)
$$
$$
[A^\mu_k, A^\nu_{k'}] = [A^{\dagger \mu}_k, A^{\dagger \nu}_{k'}] = 0 \, . \eqno(3.21)
$$

In order to attain these results it was necessary to relate the variables by the equations
$$
A^\beta_k = - \frac{i}{8 \pi^2} \int dk_0 k \sqrt{4 \pi^3(k_0^2 - k^2)} \times
$$
$$
\times \left\{\mathcal{Q}^\beta_k \left[ \delta(k_0^2 - k^2) + \frac{i}{\pi} \frac{1}{k_0^2 - k^2}\right] - \mathcal{P}^\beta_k \left[ \delta(k_0^2 - k^2) - \frac{i}{\pi} \frac{1}{k_0^2 - k^2}\right]\right\}\eqno(3.22)
$$
and
$$
A^{\dagger \beta}_{-k} =  \frac{i}{8 \pi^2} \int dk_0 k \sqrt{4 \pi^3(k_0^2 - k^2)} \times
$$
$$
\times \left\{\mathcal{Q}^\beta_k \left[ \delta(k_0^2 - k^2) - \frac{i}{\pi} \frac{1}{k_0^2 - k^2}\right] - \mathcal{P}^\beta_{-k} \left[ \delta(k_0^2 - k^2) + \frac{i}{\pi} \frac{1}{k_0^2 - k^2}\right]\right\}\, . \eqno(3.23)
$$

But these relations are not sufficient to obtain the generators in the new field variables as in Balescu \& Kotera \cite{bk}:
$$
P^0 = \sum_{j=1}^N \left\{ e_j \mathcal{A}^0(\mathfrak{q}_j) + \sqrt{[{\bf p}_j - e_j {\bf A}(\mathfrak{q}_j)]^2 + m_j^2}\right\} - 4 \pi^2 \int d^3k A^\lambda_k A^\dagger_{k \lambda} \, , \eqno(3.24)
$$
$$
{\bf P} = (P^1, P^2, P^3) = \sum_{j=1}^N {\bf p}_j - 4 \pi^2 \int \frac{d^3k}{k} {\bf k} A^\lambda_k A^\dagger_{k \lambda} \, , \eqno(3.25)
$$
$$
{\bf J} = (M^{23}, M^{31}, M^{12}) = \sum_{j=1}^N {\bf q}_j \times {\bf p}_j
$$
$$
- 4 \pi^2 \int \frac{d^3k}{k} A^{\dagger \mu}_k \left( {\bf k} \times \frac{\partial}{\partial {\bf k}}\right) A^\mu_k + {\bf A}_k \times {\bf A}^\dagger_k \, , \eqno(3.26)
$$
and
$$
{\bf K} = (M^{10}, M^{20}, M^{30}) = \sum_{j=1}^N {\bf q}_j \left\{ e_j \mathcal{A}^0(\mathfrak{q}_j) + \sqrt{[{\bf p}_j - e_j {\bf A}(\mathfrak{q}_j)]^2 + m_j^2}\right\} +
$$
$$
+ 4 \pi^2 i \int \frac{d^3k}{k} \left\{ A^\dagger_{k \mu} k \frac{\partial}{\partial {\bf q}} A^\mu_k - {\bf A}_k A^{\dagger 0}_k + {\bf A}^\dagger_k A^0_k \right\} \, , \eqno(3.27)
$$
where
$$
{\bf A}_k = (A^1_k, A^2_k, A^3_k) \, . \eqno(3.28)
$$

We found that in order to be consistent with the restrictions (3.22) and (3.23), the necessary relations for transforming our generators to the form (3.24-28) is possible in two different forms

$$
\mathcal{Q}^\mu_k = \frac{\sqrt{4\pi^3 (k_0^2 - k^2)}}{k} \times
$$
$$
\times \left\{\frac{1}{2} A^{\dagger \mu}_{-k} \delta_+(k_0 + k) + \frac{3}{4}A^\mu_k \delta_-(k_0+k) -\frac{1}{4}A^\mu_k \delta_+(k_0 -k) \right\} \, , \eqno(3.29)
$$
$$
\mathcal{P}^\mu_k = \frac{\sqrt{4\pi^3 (k_0^2 - k^2)}}{k} \times
$$
$$
\times \left\{ -\frac{1}{4} A^\mu_{-k} \delta_+(k_0+k) + \frac{1}{2} A^{\dagger \mu}_k \delta_+(k_0-k) + \frac{3}{4} A^\mu_{-k} \delta_-(k_0-k)\right\} \, . \eqno(3.30)
$$

Or
$$
\mathcal{Q}^\mu_k = \frac{\sqrt{4\pi^3 (k_0^2 - k^2)}}{k} \times
$$
$$
\times \left\{ -\frac{1}{4} A^{\dagger \mu}_{-k} \delta_+(k_0+k) + \frac{1}{2} A^\mu_k \delta_+(k_0-k) + \frac{3}{4} A^{\dagger \mu}_{-k} \delta_-(k_0-k)\right\} \, , \eqno(3.31)
$$
$$
\mathcal{P}^\mu_k = \frac{\sqrt{4\pi^3 (k_0^2 - k^2)}}{k} \times
$$
$$
\times \left\{\frac{1}{2} A^\mu_{-k} \delta_+(k_0 + k) + \frac{3}{4}A^{\dagger \mu}_k \delta_-(k_0+k) -\frac{1}{4}A^{\dagger \mu}_k \delta_+(k_0 -k) \right\} \, . \eqno(3.32)
$$

These two possibilities are related in simple form as is evident by inspection.

\section*{Concluding remarks}
We presented a Hamiltonian formalism to describe the dynamics of particles and field.

Its particularities was: an explicit tensorial covariance and a particular case of the generalized Hamiltonian dynamics elaborated by Dirac \cite{d1}, \cite{d3}. It contains actually an undetermined Lagrange multiplier for each particle.

Afterward, by a canonical transformation, we perform an arbitrary kinematical Lorentz transformation. We added a constraint to define the parametrization of the trajectories of the particles in space-time, by an arbitrary family of surfaces and in this fashion we were able to express the new Hamiltonian as a first degree polynomial in the ten translations and rotation velocities of the Lorentz transformation. The coefficients of this polynomial are the ten Dirac infinitesimal generators for the representation of the Poincar\'e group by canonical transformations \cite{d2}. The Lagrange multipliers do not appear in this transformed formalism.

In other hand, the new Hamiltonian appears as a Routh function i.e. is a Lagrangian with respect to the the variables that define the Lorentz transformation. We showed that the fundamental equations of the Lie algebra associated to the Lorentz group for the ten Dirac's generators are a consequence of the compatibility between the Lagrangian and Hamiltonian formulations associated to this Routh function.

We thus succeeded in deriving the explicit connection between two Dirac formalism \cite{d1} and \cite{d2}. The apparent difference is reconciled.

Moreover, the method here employed enables us to find the general expressions for the generators, valid for any constraint (or with any form of the dynamics, in Dirac's terminology), whereas Dirac's method uses a separated and different treatment for each form.

We found a generalized version (with any form of dynamics) for the Currie covariance conditions \cite{cu}, \cite{cj}, for the trajectories of the particles. The Currie conditions were originally derived only for the instant form of parametrization discussed in the last section of this paper. For this generalized version of Currie's conditions we found the fact stressed by Balescu and Kotera \cite{bk} that Currie's no interaction theorem \cite{cu} does not apply because of the introduction of the canonical field variables.

Finally we studied the constraint of parametrization that give us the usual Hamiltonian formulation for the system (Heitler \cite{he}). The non-trivial aspect of this case was to derive the explicit transformation between our field variables and the more usual 3-dimensional Fourier variables.

\section*{Acknowledgments}

We thank Professor I. Prigogine for the interest he took in this problem and for their kind hospitality in Brussels, where the major portion of this work was accomplished.

The author is also very much indebted to Prof. R. Balescu for suggesting this problem and for many stimulating discussions.

{\large \bf Notes }

\

This new text was totally rewritten by the author.

{\sl L'Acad\'emie Royal de Belgique} give an electronic permission for reproducing this material but did not be aware of the changes introduced.

The author modified the used fonts in order to simplify the work.

Minor print mistakes were corrected.

The last four equations used the distributions
$$
\delta_\pm(x) \equiv \delta(x) \pm \frac{i}{\pi \; x} \, .
$$

The subindex $k$ in the field coordinates $\mathcal{Q}^\mu_k$, $A^{\dagger \mu}_k$ denotes the 4-vector $k^\mu$.

Quantity $k$ in equations (3.17-28) denotes the magnitude of the 3-vector $\bf k$
$$
k \equiv |{\bf k}|.
$$

\end{document}